\documentclass[fleqn,usenatbib]{mnras}
\usepackage{newtxtext,newtxmath}
\usepackage[T1]{fontenc}
\usepackage{ae,aecompl}
\usepackage{anyfontsize}

\usepackage{booktabs}

\usepackage{graphicx}	
\usepackage{amsmath}	
\usepackage{amssymb}	
\usepackage{hyperref}
\usepackage[pdfpagelabels=false]{hyperref}	
\hypersetup{colorlinks=true,
            linkcolor=blue,
            citecolor=blue,
            filecolor=blue,
            urlcolor=blue,}

\hypersetup{pageanchor=false}

\title[Rms amplitude of kHz QPOs in 4U 1636\(-\)53]{The amplitude of the kilohertz quasi-periodic oscillations  \\ in 4U~1636\(-\)53 in the frequency-energy space}

\author[E. M. Ribeiro et al.]{
 Evandro M. Ribeiro\(^{1}\)\thanks{E-mail: ribeiro@astro.rug.nl},
 Mariano M\'{e}ndez\(^{1}\), Marcio G. B. de Avellar \(^{2}\),
\newauthor
Guobao Zhang \(^{3,\ 4}\) and Konstantinos Karpouzas \(^{1}\)
\\
\(^{1}\) Kapteyn Astronomical Institute, University of Groningen, P.O. BOX 800, 9700 AV Groningen, The Netherlands\\
\(^{2}\) Instituto Tecnol\'ogico de Aeron\'autica, ITA, Brazil\\
\(^{3}\) Yunnan Observatories, Chinese Academy of Sciences (CAS), Kunming 650216, P.R. China \\
\(^{4}\) Key Laboratory for the Structure and Evolution of Celestial Objects, CAS, Kunming 650216, P.R. China \\
}

\date{Accepted XXX.\@ Received YYY;\@ in original form ZZZ}

\pubyear{2019}

\begin{document}\label{firstpage}
\pagerange{\pageref{firstpage}--\pageref{lastpage}}
\maketitle

\begin{abstract}
We present for the neutron-star low-mass X-ray binary 4U~1636\(-\)53, and for the first time for any source of kilohertz quasi-periodic oscillations (kHz QPOs), the two-dimensional behaviour of the fractional rms amplitude of the kHz QPOs in the parameter space defined by QPO frequency and photon energy. We find that the rms amplitude of the lower kHz QPO increases with energy up to \(\sim12\)~keV and then decreases at higher energies, while the rms amplitude of the upper kHz QPO either continues increasing or levels off at high energies. The rms amplitude of the lower kHz QPO increases and then decreases with frequency, peaking at \(\sim 760\)~Hz, while the amplitude of the upper kHz QPO decreases with frequency, with a local maximum at around \(\sim 770\)~Hz, and is consistent with becoming zero at the same QPO frequency, \(\sim1400\)~Hz, in all energy bands, thus constraining the neutron-star mass at $M_{NS} \leq 1.6 M_{\odot}$, under the assumption that this QPO reflects the Keplerian frequency at the inner edge of the accretion disc. We show that the slope of the rms energy spectrum is connected to the changing properties of the kHz QPOs in different energy bands as its frequencies change. Finally, we discuss a possible mechanism responsible for the radiative properties of the kHz QPOs and, based on a model in which the QPO arises from oscillations in a Comptonising cloud of hot electrons, we show that the properties of the kHz QPOs can constrain the thermodynamic properties of the inner accretion flow.
\end{abstract}


\section{Introduction}\label{sec:intro}

Twenty years after the discovery of kilohertz quasi-periodic oscillations ~\citep[kHz QPOs,][]{Strohmayer1996, VanderKlis1996} in neutron-star low-mass X-ray binaries (NS-LMXB), a satisfactory explanation of the origin of these QPOs remains a challenge~\citep{Gilfanov2003, Mendez2006, DeAvellar2013, Barret2013, Peille2015, Wang2016, Cackett2016, Troyer2018}.
Both the timing and spectral properties of these high-frequency oscillations are likely driven by general relativistic effects~\citep{Stella1998, Stella1999, Kato2005}, the neutron-star mass and radius \citep{Miller1998}, and the physical processes taking place in the inner regions of the accretion flow~\citep{Lee1997, Lee2001, Kumar2014}.

The kHz QPOs usually appear as a pair of peaks in the X-ray power density
spectrum (PDS) of NS-LMXB. The two components of the pair of kHz QPOs are called the lower and upper kHz QPO, respectively, according to their frequency. In those cases in which a single QPO is present in the PDS of an observation, the QPO can be identified by other properties such as their amplitude~\citep{DiSalvo2001, Mendez2001}, quality factor \(Q = \nu_0/\text{FWHM}\), where \(\nu_0\) is the QPO frequency and FWHM is the full width at half maximum of the profile of the QPO \citep{Barret2006a}, the relation between the QPO frequencies and the spectral state of the NS-LMXB~\citep{Mendez1998a}, or the time lags of the QPOs~\citep{Barret2013, DeAvellar2013, Peille2015, DeAvellar2016, Troyer2018}.
For instance, the upper kHz QPO has usually higher fractional rms amplitude and lower quality factor than the lower kHz QPO~\citep[e.g.,][]{DiSalvo2001, vanStraaten2002, Barret2006a, Mendez2006, Altamirano2008a, Troyer2018} and is observed across different spectral states, whereas the lower kHz QPO is only observed in the intermediate state~\citep[e.g.][]{Mendez1998a, Zhang2016} .

Several models have been put forward in order to explain the frequency of the kHz QPOs in LMXB, among those are the sonic-point model~\citep{Miller1998}, the relativistic precession model~\citep{Stella1998}, the two-oscillator model~\citep{Osherovich1999, Titarchuk2003}, the relativistic resonant model~\citep{Kluzniak2001}, the deformed-disc oscillation model~\citep{Kato2009, Mukhopadhyay2009}, and magneto-hydrodynamic models~\citep{Zhang2004, Li2005,  Erkut2008, Shi2009, DeAvellar2018}. So far none of these models can satisfactorily explain all of the observed properties of the kHz QPOs \citep[e.g.][]{Lin2011, Wang2016}.

Another approach to understand the origin of the kHz QPOs is to investigate the relation between the properties of the QPOs and the X-ray flux and spectra and to identify the radiative mechanism responsible for the QPOs.
It is known that the frequencies of these oscillations depend on the spectral state of the source, which can be parametrised by the position of the source in the colour-colour diagram \citep{Mendez1998a, Zhang2016, Ribeiro2017b}. 
On the other hand, the fractional rms amplitude of the kHz QPOs increases with photon energy up to 10~keV \citep{Berger1996a, Zhang1996, Wijnands1997, Mendez2001, Gilfanov2003}, and it may decrease at energies above \(\sim~10\)~keV \citep{Mukherjee2012}.
The time-averaged X-ray spectrum of NS LMXBs is composed by a soft thermal component due to the neutron-star surface/boundary layer and the accretion disc and a hard component due to the inverse Compton scattering of soft thermal photons in a corona consisting of energetic electrons \citep{Barret2001, Lin2007}.
It is worth noticing that at energies above \(\sim  10\)~keV the contribution of the soft thermal component to the total emission becomes insignificant, implying that the radiative mechanism responsible for the amplitude of the kHz QPOs must take place in the hard Comptonising component \citep{Gilfanov2003, Mendez2006}.

In previous studies of this source,~\cite{Zhang2016} explored the connection between the presence of the lower and upper kHz QPO and the spectral state of the source, and~\cite{Ribeiro2017b} studied the relation between the QPO properties and the spectral parameters.    
In this work we study the evolution of the fractional rms amplitude of the kHz QPOs in 4U 1636\(-\)53 both as a function of the QPO frequency and energy.

In this paper we combine data from the spectral and timing domain to reveal new information on the properties of the kHz QPOs in the NS-LMXB 4U 1636\(-\)53.
This source shows a recurring \(\sim 40\)-day cycle of spectral transitions~\citep{Belloni2007}, making it an excellent target to study the changes of the properties of the kHz QPOs.
4U 1636\(-\)53 was observed 1576 times with the Rossi X-ray Timing Explorer (RXTE) satellite, providing an outstanding data set to carry out detailed analysis of the kHz QPO.

In \S 2 we explain the methods, in \S 3 we show our results and in \S 4 we discuss our findings and present our conclusions about the results of our novel approach.

\section{Methods}\label{sec:methods}

From the 1576 RXTE observations of the NS-LMXB 4U~1636\(-\)53, we selected those that were carried out in Event mode with a time resolution of at least 125 \( \mu \)s.
We extracted power spectra of the selected observations using all PCA energy channels over 16-s segments with a Nyquist frequency of 2048 Hz. We discarded those 16-s segments in which there were telemetry drop-outs or an X-ray burst was present in the light curve, and we averaged the remaining 16-seconds power spectra to produce a single power spectrum per observation.
We searched the average power spectrum of each observation for kHz QPOs at frequencies larger than \(200\) Hz.
The criteria used to identify kHz QPOs were the same as described in~\cite{Sanna2012}, \cite{ Zhang2016} and~\cite{Ribeiro2017b}: we fitted the average power spectra with a model consisting of a constant component to represent the power produced by the Poissonian nature of the light curve, plus one or two Lorentzians to represent the kHz QPOs, and we accepted as QPOs those peaks in the averaged power spectrum of an observation where the ratio between the normalisation of the Lorentzian function and its negative 1\( \sigma \) error was larger than 3, and the quality factor \(Q\) was larger than 2. We ended up with \(580\) observations with at least one kHz QPO detected.

\begin{figure}
    \centering
 	\includegraphics[width=0.9\columnwidth]{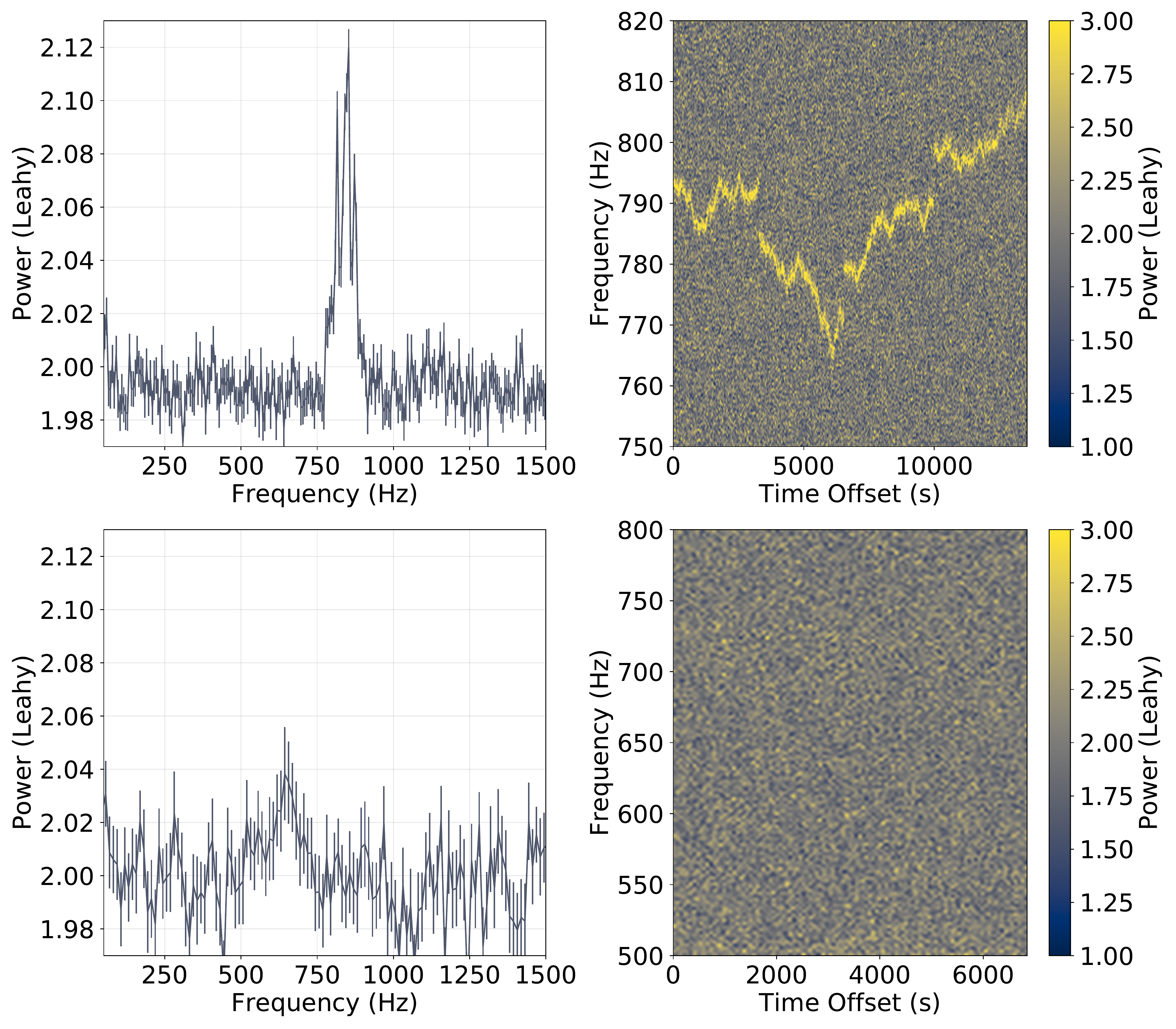}
     \caption{\label{fig:dynspec} Example of two observations with a kHz QPO in 4U~1636\(-\)53. The left column shows the averaged power-spectrum of the observation, and the right column shows the corresponding dynamical power spectrum.
     \textbf{Top row}: good quality observation, in which we are able to trace the QPO in each non-overlapping segment of 16-s. The multiple peak profile of the QPO in the left panel is due to the change of the QPO frequency during the observation. The small gaps in the dynamical power spectrum, apparent as small jumps of the frequency of the QPO at around 2500s, 7000s and 10000s, come from the selection of \textit{Good Time Intervals}.
     \textbf{Bottom row}: worst case scenario, in which we are not able to identify the QPO in the dynamical power spectrum, but we are still able to detect the QPO in the averaged power spectrum. In this case we assigned a single frequency to the QPO for the entire observation.
     }
\end{figure}

Once we identified the observations with kHz QPOs, we created dynamical power spectra~\citep{Berger1996a} using the non-overlapping 16-s segments in order to trace the frequency evolution of the QPOs during each observation.
In the best cases we were able to trace the frequency evolution down to 16~s, whereas in the worst cases we assigned the frequency measured from the averaged PDS for each of the detected QPOs for an entire observation. \autoref{fig:dynspec} shows two examples of dynamical power spectra in the best and worst case scenarios. For the intermediate cases we combined as many non-overlapping 16-s segments as necessary to identify the QPO in the dynamical power spectra.
At the end of this procedure we were able to assign at least one QPO frequency to each 16-seconds segment of an observation\footnote{In the worst case scenario explained above, we assigned the same frequency to all 16-s intervals of an observation.}.
If there were two simultaneous kHz QPOs in an observation, we assigned two frequencies to each segment of that observation, corresponding to the lower and upper kHz QPOs, respectively. For observations with only one kHz QPO we used the colour-colour diagram to identify the QPO as the lower or the upper kHz QPO as done by \cite{Ribeiro2017b}.

We divided the frequency range covered by the QPOs into 9 frequency intervals for the upper kHz QPO and 11 frequency intervals for the lower kHz QPO, as shown in \autoref{tab:selections}.
This selection is the same used by~\cite{Ribeiro2017b} and was chosen to achieve a compromise between having a sufficient number of observations with QPOs at each of the frequency intervals, and having enough frequency intervals for a meaningful analysis.
We then selected and combined all 16-s power spectra with an assigned QPO frequency in each of those intervals, separately for each QPO.
Note that this is different from the technique used in~\cite{Zhang2016} and~\cite{Ribeiro2017b}, where each observation was assigned only one averaged QPO frequency (for each kHz QPO detected), which corresponds to the approach of the worst case scenario in this work, illustrated in the bottom row of \autoref{fig:dynspec}.

\begin{table}
  \centering
	\caption{\label{tab:selections}
    Overview of the QPO frequency intervals used to combine the different power spectra of 4U~1636\(-\)53.
    The uncertainties of the rms fractional amplitude and QPO frequency represent the 1\(\sigma \) confidence interval propagated from the best fitted Lorentzian and the modelled background count-rate, as explained in the text.
    }
	\begin{tabular}{lccc} 
    \toprule
         Frequency & \multicolumn{1}{l}{Number of}& \multicolumn{1}{l}{Average} & \\
         interval (Hz) & 16-s segments & frequency (Hz)   & rms (\%) \\
    \midrule
         \multicolumn{4}{l}{Lower kHz QPO} \\
    \midrule
         \(470\)--\(590\) & \(1859 \) & \(574.1 \pm 2.6\)  & \(4.39 \pm 0.40\) \\
         \(590\)--\(620\) & \(2141\)  & \(611.1 \pm 1.2\)  & \(4.55 \pm 0.32\) \\
         \(620\)--\(670\) & \(8009\)  & \(651.9 \pm 0.8\)  & \(6.62 \pm 0.38\) \\
         \(670\)--\(715\) & \(6302 \) & \(698.3 \pm 0.5\)  & \(7.36 \pm 0.21\) \\
         \(715\)--\(750\) & \(6385\)  & \(732.1 \pm 0.4\)  & \(8.08 \pm 0.27\) \\
         \(750\)--\(790\) & \(6862\)  & \(765.2 \pm 0.4\)  & \(8.22 \pm 0.32\) \\
         \(790\)--\(820\) & \(7573\)  & \(809.1 \pm 0.2\)  & \(7.86 \pm 0.19\) \\
         \(820\)--\(850\) & \(9569\)  & \(838.7 \pm 0.2\)  & \(7.55 \pm 0.34\) \\
         \(850\)--\(880\) & \(10738\) & \(864.2 \pm 0.1\)  & \(7.15 \pm 0.21\) \\
         \(880\)--\(910\) & \(11919\) & \(894.4 \pm 0.2\)  & \(5.42 \pm 0.19\) \\
         \(910\)--\(975\) & \(8422\)  & \(917.8 \pm 0.4\)  & \(3.88 \pm 0.10\) \\
    \midrule
        \multicolumn{4}{l}{Upper kHz QPO} \\
    \midrule
        \(440\)--\(540\)   & \(4033\) & \(486.2  \pm 6.4\)  & \(13.7 \pm 0.8\) \\
        \(540\)--\(650\)   & \(3872\) & \(604.4  \pm 2.9\)  & \(11.9 \pm 0.5\) \\
        \(650\)--\(750\)   & \(6911\) & \(718.7  \pm 2.2\)  & \(12.4 \pm 0.5\) \\
        \(750\)--\(810\)   & \(7662\) & \(780.0  \pm 1.5\)  & \(11.3 \pm 0.4\) \\
        \(810\)--\(870\)   & \(5258\) & \(838.5  \pm 1.5\)  & \(10.0 \pm 0.4\) \\
        \(870\)--\(930\)   & \(4669\) & \(892.3  \pm 2.1\)  & \(8.6  \pm 0.3\) \\
        \(930\)--\(1025\)  & \(1316\) & \(947.0  \pm 3.8\)  & \(5.8  \pm 0.4\) \\
        \(1025\)--\(1165\) & \(382\)  & \(1147.1 \pm 17.1\) & \(3.0  \pm 1.3\) \\
        \(1165\)--\(1250\) & \(4087\) & \(1223.1 \pm 2.8\)  & \(2.3  \pm 0.1\) \\
    \bottomrule
	\end{tabular}
\end{table}

For each observation with a kHz QPO, we computed PDS at various energy bands. The chosen energy bands are shown in~\autoref{tab:selections_en}. This energy selection is the same one used by~\cite{DeAvellar2016}.
We took into account the slow drift in the energy-to-channel relation of the PCA detectors to define the boundaries of the energy bands in the different RXTE/PCA gain epochs \citep{DeAvellar2013}.
We then measured the rms amplitude of each QPO in the averaged PDS for each of the selected frequency intervals in the full energy band (nominally \(2\)--\(60\) keV), and at the selected energy bands as follows:

\begin{table}
  \centering
	\caption{\label{tab:selections_en} Overview of energy bands used to select individual power spectra.
    The channel boundaries of each band were adapted for each RXTE/PCA epoch to correspond approximately to the same energy band.
    The uncertainties of the rms fractional amplitude represent the 1\(\sigma \) confidence interval propagated from the best fitted Lorentzian and the modelled background count-rate, as explained in the text, we measured the rms amplitude in the power spectra after shifting the lower and upper kHz QPOs, respectively, to a single frequency using the shift-and-add technique.  }
	\begin{tabular}{lccccc} 
    \toprule
     Average      & \multicolumn{3}{l}{Channel selection by Epoch} & \multicolumn{2}{c}{rms (\%)}\\
		 Energy & 3rd       & 4th       & 5th     & Lower & Upper \\
		 (keV) & Epoch & Epoch & Epoch & kHz QPO & kHz QPO \\
     \midrule
		\(4.2\)     & \(08\)--\(12\) & \(07\)--\(10\) & \(07\)--\(11\) &
		\(4.3 \pm 0.4\) & \(5.4 \pm 0.3\) \\
		\(6.0\)     & \(13\)--\(17\) & \(11\)--\(05\) & \(12\)--\(15\) &
		\(6.0 \pm 0.2\) & \(6.4 \pm 0.5\) \\
		\(8.0\)     & \(18\)--\(23\) & \(16\)--\(21\) & \(16\)--\(21\) &
		\(8.2 \pm 0.3\) & \(9.3 \pm 0.5\) \\
        \(10.2\)    & \(24\)--\(29\) & \(22\)--\(25\) & \(22\)--\(25\) &
        \(9.5 \pm 0.6\) & \(13.1 \pm 1.3\) \\
      	\(12.7\)    & \(30\)--\(41\) & \(26\)--\(35\) & \(26\)--\(35\) &
      	\(11.0 \pm 0.8\) & \(13.1 \pm 1.1\) \\
      	\(16.3\)    & \(42\)--\(46\) & \(36\)--\(39\) & \(36\)--\(41\) &
      	\(9.8 \pm 1.2\) & \(16.2\ ^{*} \)\\
      	\(18.9\)    & \(47\)--\(55\) & \(40\)--\(46\) & \(42\)--\(49\) &
      	\(7.2 \pm 1.3\) & \( 19.1\ ^{*} \)\\
    \bottomrule
        \multicolumn{4}{l}{\(^{*}\) Upper limit at \(95 \% \) confidence. }
   \end{tabular}
\end{table}

We fitted the averaged PDS in the frequency range of 200--1500~Hz in each energy band and for each QPO frequency interval with a constant, to represent the Poisson noise, and 1 or 2 Lorentzians to represent the kHz QPOs.
We binned the power-spectra by a factor of 50, which yielded a frequency resolution of \(\sim 3\) Hz.
At first we let all the parameters of the QPOs free between energy bands. When we were not able to detect a significant QPO at a specific energy band, we fixed the frequency and width of the QPO to be the same values as those in the full-energy band to calculate upper limits.

We used the integrated power of the best-fitting Lorentzian, $P$, to represent the total power of the QPO. We then calculated the rms amplitude in percent units as:

\begin{equation}
rms = \sqrt{\frac{P}{S + B}} \cdot \left(\frac{S+B}{S}\right) \cdot 100 \; ,
\end{equation}

where $S$ is the source count rate and $B$ is the background count rate. We calculated the source count rate as the difference between the total observed count rate, $C$, and the background count rate.

To estimate the background count rate we used the \textit{Ftool} \texttt{pcabackest} based on the \texttt{Standard2} light curves of each observation, for each of the energy bands investigated.
We measured, for each of the frequency intervals, the average and the peak-to-peak variation of the background count rate of the combined observations with QPO frequency within that interval.

Using the procedure above we obtained the rms amplitude of both kHz QPOs in each energy band and frequency interval.
We used \texttt{PyXspec} version \(2.0.1\) \citep{Arnaud1996} to fit the rms-vs-energy and rms-vs-frequency relations with simple analytical models.
The models fitted have no physical motivation, and some of the fits are statistically unacceptable, but they reveal the quantitative differences between the different relations.

In all the fit procedures presented in the next section we used the 1-\(\sigma\) error to represent the uncertainties of the fitted variable.
On the occasions where the fitted Lorentzian power was not significantly different from zero but still positive, we used the calculated values of the fractional rms amplitude as valid measurements during the fit procedures.
In some cases the best-fitting Lorentzian yielded a negative integrated power at the expected QPO frequency in relation to the Poisson level; for those cases we used zero as the value of the rms amplitude and the 95\% confidence upper limit as the error bar for the fit routines.
In both cases we plot the 95\% confidence upper limit to provide the correct visualisation, and in the Tables we give both the measurements and the upper limits.


\section{Results}\label{sec:results}

In this section we show the results of the rms amplitude of both kHz QPOs in the 2-dimensional space of QPO frequency and photon energy. In \autoref{sec:marginals} we show the rms amplitude marginalised over, respectively, QPO frequency and energy, which we call the marginal distributions of the rms amplitude. In \autoref{sec:conditionals} we show the results of, respectively, the rms amplitude vs. QPO frequency given the photon energy and rms amplitude vs. energy given the QPO frequency, which we call the conditional distributions of the rms amplitude. Finally, in \autoref{sec:2dimensions} we show the joint distribution of the rms amplitude vs QPO frequency and photon energy.

\subsection{The marginal distributions of rms amplitude vs. QPO frequency and photon energy}\label{sec:marginals}

\begin{figure}
    \centering
 	\includegraphics[width=0.9\columnwidth]{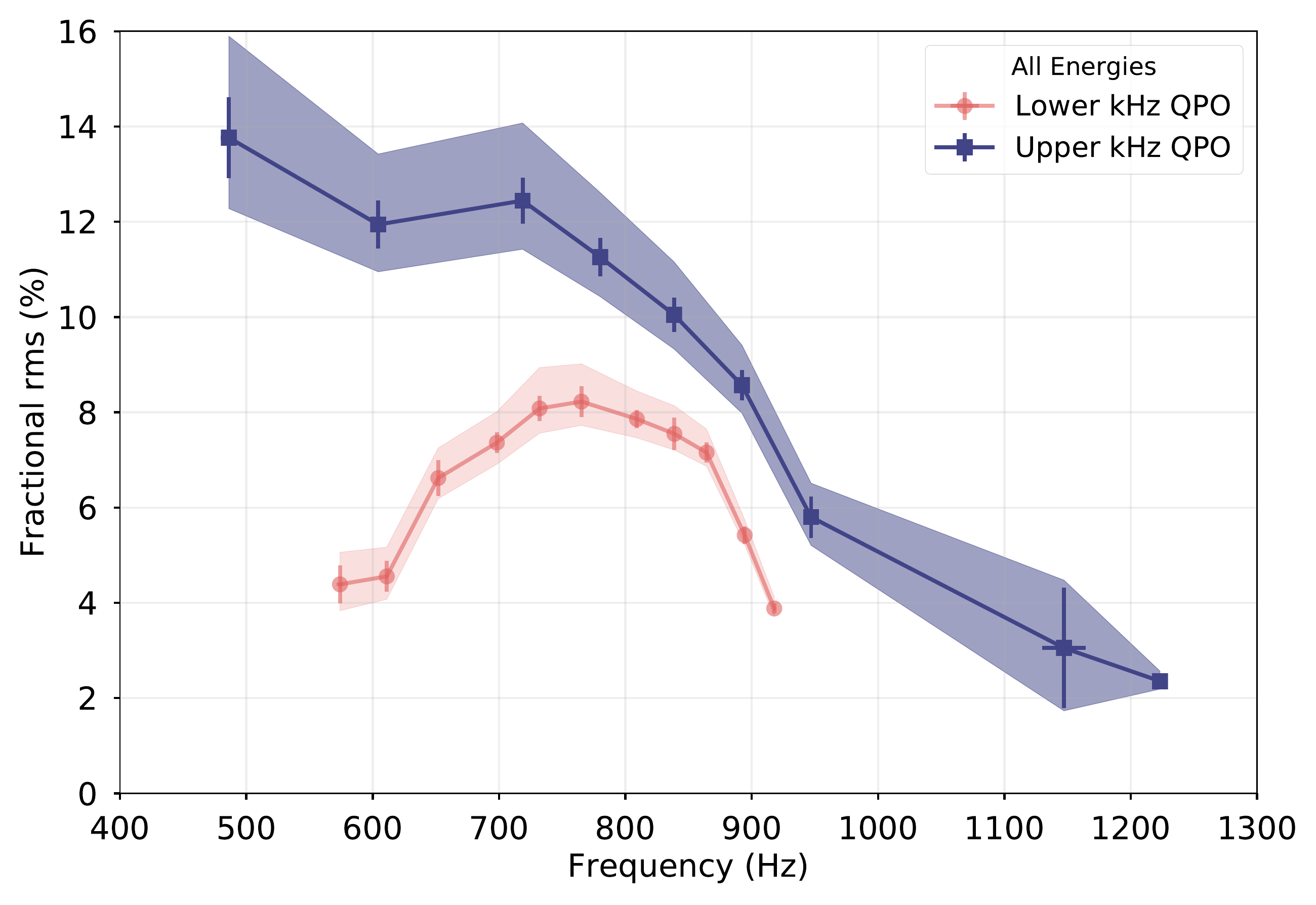}
     \caption{\label{fig:rms_freq_full} The marginal distribution of the rms amplitude of the kHz QPOs of 4U~1636\(-\)53 as a function of QPO frequency, averaged over the full PCA energy band. The lower kHz QPO is shown in light red and the upper kHz QPO in dark blue. The shaded areas represent the range of rms values assuming a background count rate between zero and two times larger than the maximum value given by the \texttt{pcabackest} tool, including the statistical errors in those cases.
     (A colour version of this figure is available in the on-line
      version of the paper).
      }
\end{figure}

In \autoref{fig:rms_freq_full} we show the marginal distribution of the rms amplitude of the lower and upper kHz QPO in 4U~1636\(-\)53 in the full energy band of the PCA detectors, nominally \(2\)--\(60\)~keV, as a function of frequency.
We used the shift-and-add technique \citep{Mendez1997} to average the power spectra with different QPO frequencies and obtain the marginal distribution of rms amplitude as a function of photon energy (see below). For the conditional distributions (\autoref{sec:conditionals}) of the rms fractional amplitude as a function of frequency and energy we did not use the shift-and-add technique since the frequency of the QPO does not change much within the chosen frequency intervals and the averaged PDS can be fitted directly with a Lorentzian in order to obtain the QPO rms amplitude.

It is apparent from \autoref{fig:rms_freq_full} that in 4U~1636$-$53 the relation between amplitude and frequency of the lower kHz QPO is different than that of the upper kHz QPO.
The rms amplitude of the lower kHz QPO increases with frequency from \(\sim 4 \) \% at \(570\) Hz up to \(\sim 8\% \) at \(750\) Hz, where it peaks, and then decreases back to \(\sim 4\% \) as the frequency increases up to \(\sim 920\) Hz.
On the contrary, the rms amplitude of the upper kHz QPO decreases from \( \sim 14\% \) down to \( \sim 2\% \) as the QPO frequency increases from \(480\) to \(1200\) Hz \citep{DiSalvo2001, Mendez2001, Barret2005b, Mendez2006}: As previously noted by \cite{Ribeiro2017b}, besides the overall decreasing trend, the rms amplitude of the upper kHz QPO shows a hump at around \(\sim 700\) Hz.

To check the effect of the modelling of the background upon our results, we estimated the change of the fractional rms amplitude if the averaged background count rate, $B$, was either zero or a factor of 2 times larger than the maximum value obtained using \texttt{pcabackest} as described in \autoref{sec:methods}.
We show the result of this test by the shaded area in \autoref{fig:rms_freq_full}.
We chose a factor of 2 to demonstrate the robustness of the rms values.
As can be seen from the Figure, the trends are not significantly affected by this exaggerated uncertainty on the background count rate~\citep[See also][]{DeAvellar2016}.

We fitted the relations in \autoref{fig:rms_freq_full} with simple analytical functions, as done by \cite{Ribeiro2017b}. For the lower kHz QPO we fitted a Gaussian function defined as:

\begin{equation}\label{eq:gauss}
    f(\nu) = N \cdot e^{-\frac{(\nu-\nu_0)^2}{2 \cdot \sigma^2}} \; ,
\end{equation}

\noindent where $\nu$ is the QPO frequency, $\nu_0$ is the centroid, $N$ is the height, and $\sigma$ is the standard deviation of the Gaussian. The best fit yielded $\nu_0 = 761 \pm 2$~Hz, $\sigma = 134 \pm 3$~Hz and $N = 8.6 \pm 0.1$~\%.

For the upper kHz QPO we fitted a model consisting of a linear function plus a Gaussian. The linear function was defined as:

\begin{equation}\label{eq:uplinear}
        f(\nu) = - s_1 \cdot (\nu - \nu_{i}) \ \text{,}
\end{equation}

\noindent where $\nu_{i}$ is the frequency at which the rms is zero, $s_1$ is the slope of the function and $\nu$ is the QPO frequency.
The best fit yielded a slope $s_1 = 0.014 \pm 0.01$~\%/Hz, an intercept $\nu_{i} = 1381 \pm 16$~Hz and a Gaussian centred at $\nu_0 = 765 \pm 17$~Hz with $\sigma = 103 \pm 21$~Hz and $N=2.6 \pm 0.5$~\%. These results are consistent with the ones presented by \cite{Ribeiro2017b}, with the small differences due to the fact that we are averaging power spectra taken  every 16-s, whereas in \cite{Ribeiro2017b} we averaged the power spectra for each observation before separating them into frequency intervals.

\begin{figure}
    \centering
 	\includegraphics[width=0.9\columnwidth]{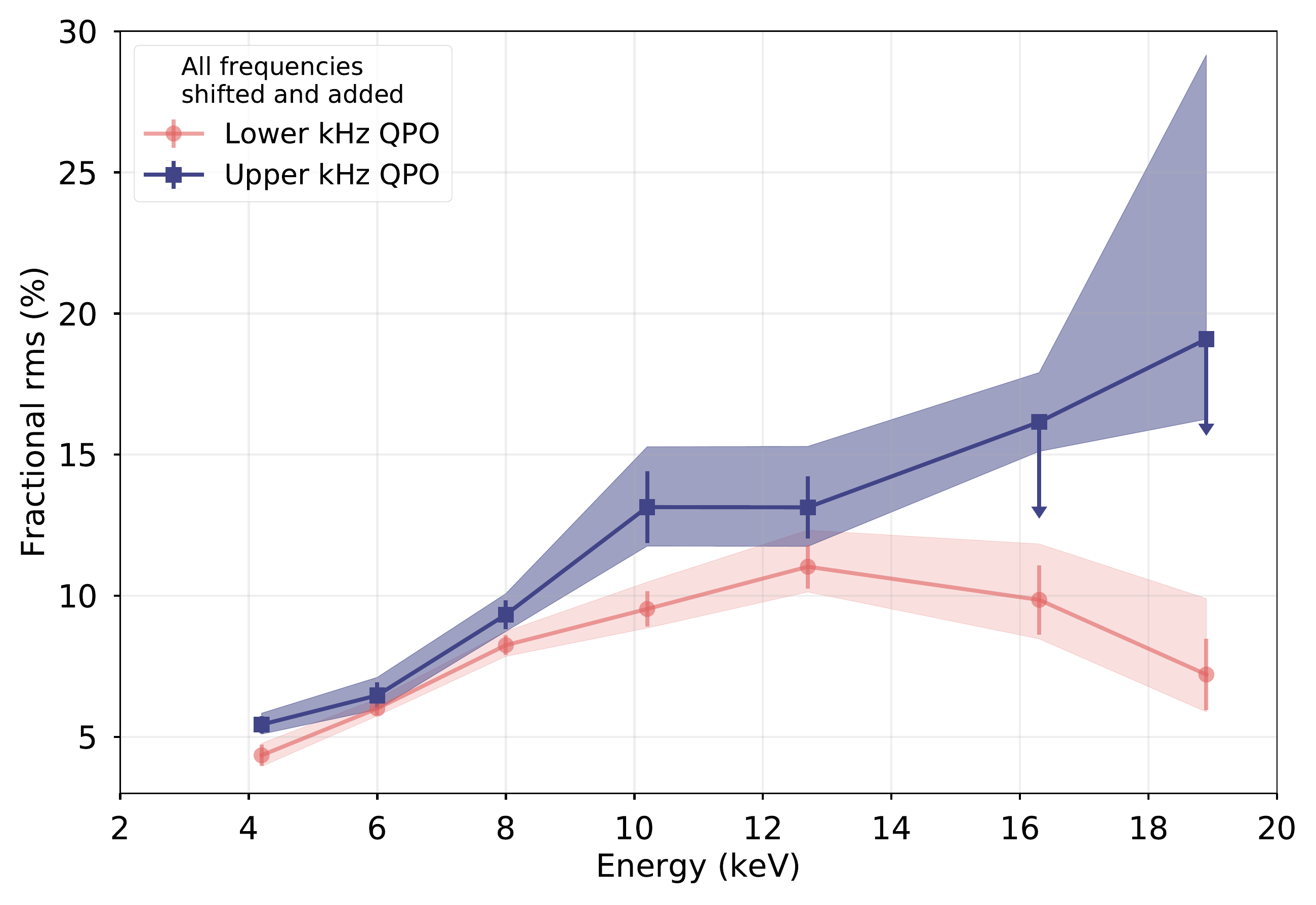}
     \caption{\label{fig:rms_allfreq} The marginal distribution of the rms amplitude of the kHz QPOs of 4U~1636\(-\)53 as a function of photon energy, averaged over all detected QPO frequencies using the shift-and-add technique \citep{Mendez1997} separately for the lower and upper kHz QPOs. The lower kHz QPO is shown in light red and the upper kHz QPO in dark blue. The shaded areas represent the range of rms values assuming a background count rate between zero and two times larger than the maximum value given by the \texttt{pcabackest} tool, including the statistical errors in those cases. When an upper limit is present the lower end of shaded area is also an upper limit.
     (A colour version of this figure is available in the on-line
      version of the paper)
     }
\end{figure}

In \autoref{fig:rms_allfreq} we show the marginal distribution of the rms amplitude of the kHz QPOs as a function of energy.
To fit the QPO in each band we used the shift-and-add technique \citep{Mendez1997} to shift the QPOs of all the observations to a common frequency. We applied this shift separately for the lower and upper kHz QPOs.
The rms amplitude of the upper kHz QPO increases from \(\sim 5\) \% to \(\sim 17\) \% as the energy increases from 4 to 19 keV. (The two last measurements are upper limits of the rms amplitude). As shown by the shaded area, at energies above 13~keV the measurements are affected by the uncertainties in the background count rate.
On the other hand, the rms amplitude of the lower kHz QPO increases  at first from \(\sim 5\) \% to \(\sim 11\) \% as the energy increases from 4 to \(\sim 12\) keV, and then decreases to \(\sim 7\) \% as the energy increases from \(\sim 12\) to 19 keV. As we can see from the shaded area limits, if we allow the background count rate to vary from zero to a factor of 2 higher than the maximum value estimated using \texttt{pcabackest}, the effect of this systematic error on the rms values does not change the observed shape of the rms as a function of energy.

We proceeded to quantify the shape of the marginal distribution of the rms amplitude vs. energy in the same manner as for the marginal distribution of rms amplitude vs. frequency, above.
We first fit the rms-vs-energy relation of the lower kHz QPO with a linear function of the type $rms(E) = s_1 \times E$, which results in a linear slope of $s_1 = 0.88 \pm 0.02$ \%/keV with a reduced chi-squared value $\chi^{2}_{\nu} = 15.2$ for 6 degrees of freedom.
We then fit the same rms-vs-energy relation with a broken-line function defined as:

\begin{equation}\label{eq:brokenline}
    rms(E)= 
\begin{cases}
    s_1\cdot E,& \text{if } E \leq E_{break}\\
    s_2\cdot E + (s_1-s_2)\cdot E_{break},  & \text{if } E \geq E_{break}
\end{cases}    
\end{equation}

\noindent where $E_{break}$ is the break energy, and $s_1$ and $s_2$ are the slopes before and after the break, respectively.
The fit yielded a reduced chi-squared of $\chi^{2}_{\nu} = 0.55$ for 4 degrees of freedom.
Compared to the fit with a linear function, the F-test probability is $6 \cdot 10^{-4}$.
The broken line model yielded a slope before the break $s_1 = 0.99 \pm 0.02$ \%/keV, slope after the break $s_2 = -0.6 \pm 0.2$ \%/keV, and a break energy $E_{break} = 11.8 \pm 0.3$ keV.
Compared to a fit with the slope after the break fixed to zero the F-test probability is $2.8 \times 10^{-2}$.

For the upper kHz QPO a linear function with slope $s=1.15 \pm 0.04$ \%/keV yielded a reduced chi-squared $\chi^{2}_{\nu} = 1.89$ for 6 degrees of freedom.
A broken line model yields a reduced chi-squared $\chi^{2}_{\nu} = 1.33$ for 4 degrees of freedom, with $s_1 = 1.18 \pm 0.04$ \%/keV, $s_2 = 2.7 \pm 2.7$ \%/keV and $E_{break} = 12.4 \pm 1.5$ keV. The F-test between the linear and a broken-line model yields a probability of $0.22$.
Since the slope after the break was not significantly different from zero, we fixed $s_2 = 0$, which yielded a reduced chi-square $\chi^{2}_{\nu} = 1.27$ for 5 degrees of freedom, the slope before the break is $s_1 = 1.18 \pm 0.04$ \%/keV and the break energy $E_{break} = 10.9 \pm 0.9$ keV.
An F-test between the broken line models with $s_2$ free to vary and with $s_2$ fixed to zero yields a probability of $0.43$.

\subsection{The conditional distributions of the rms amplitude vs. QPO frequency and energy}\label{sec:conditionals}

In this section we study the relation between rms amplitude and frequency of the QPOs in different energy bands, and of the rms amplitude of the QPOs and energy at different QPO frequencies, which we call conditional distributions.
Initially, we looked at observations that contained a QPO detectable in most energy bands and confirmed that the frequency of the kHz QPO in different bands is consistent with being the same, within the chosen frequency intervals, during the same observation. This is an important information to investigate the nature of kHz QPOs \citep{Mukherjee2012, Wang2016}.

\begin{figure*}
    \centering
 	\includegraphics[width=0.9\paperwidth]{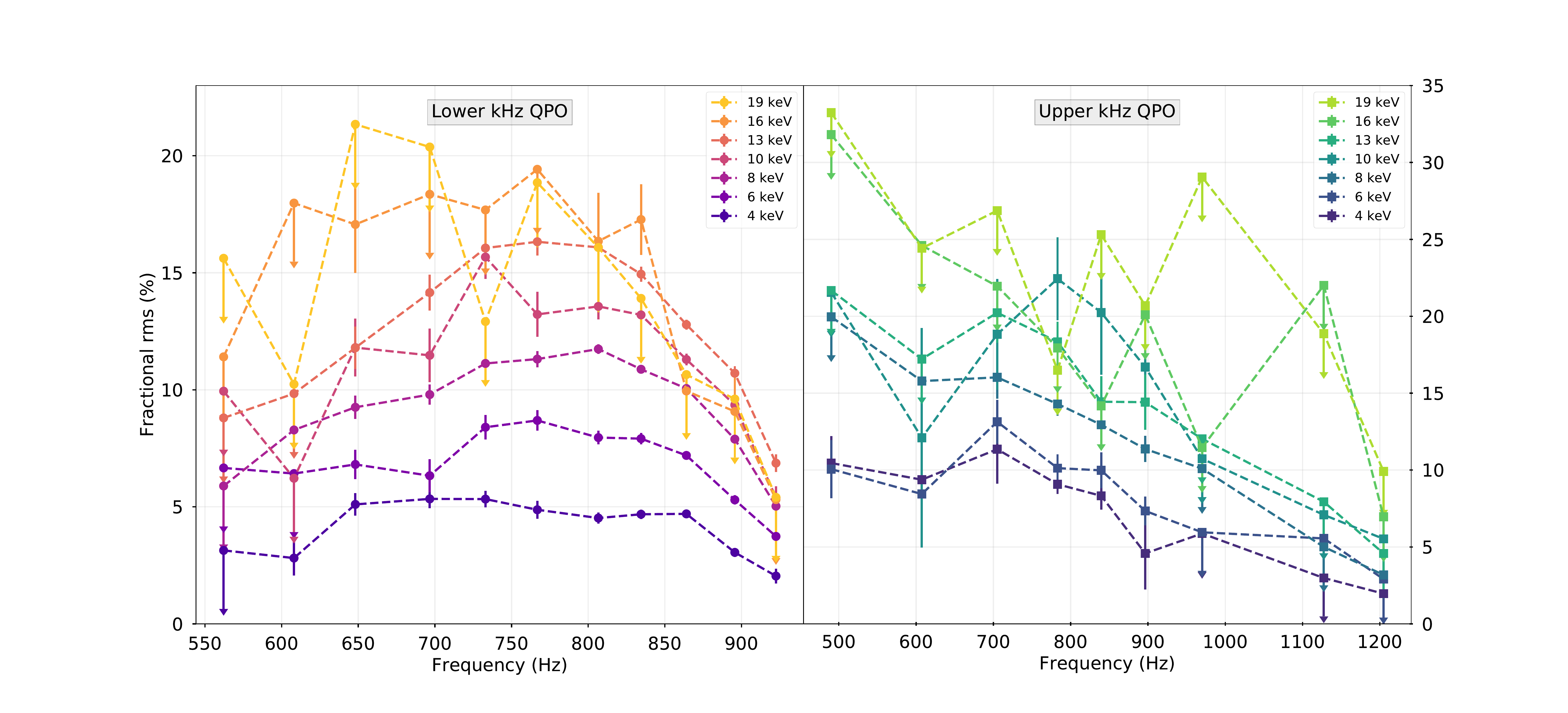}
     \caption{\label{fig:rms_freq_all} The conditional distribution of the rms amplitude of the lower and upper kHz QPOs, left and right panels, respectively, of 4U~1636\(-\)53 as a function of QPO frequency for a given energy. Each energy band is represented by a different colour.
     (A colour version of this figure is available in the on-line version of the paper).
     }
 \end{figure*}

\begin{figure}
    \centering
    \includegraphics[width=0.9\columnwidth]{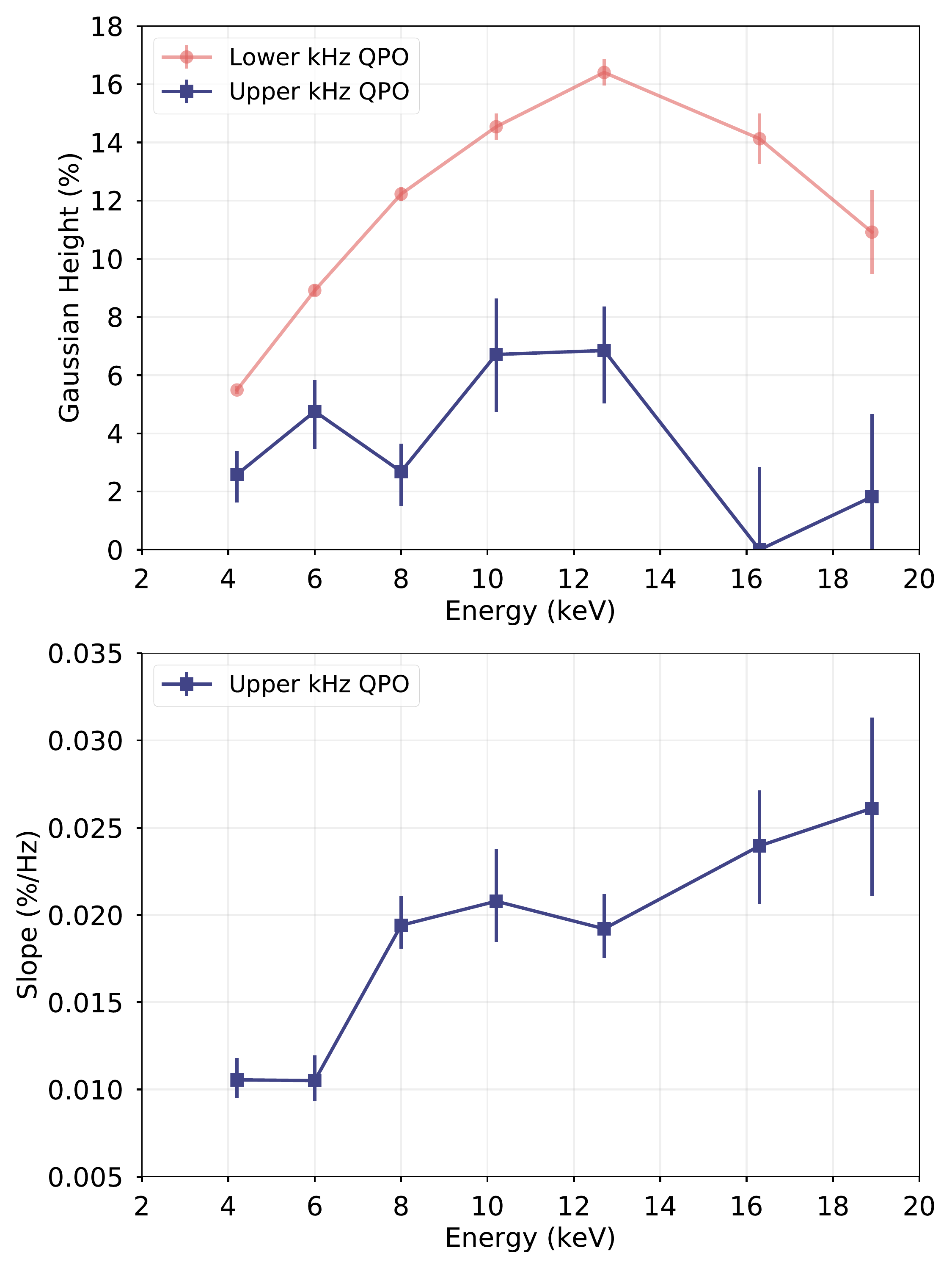}
    \caption{\label{fig:normgausian} Upper panel: The normalisation of the Gaussian from the best-fitting model to the rms amplitude of the lower kHz QPO (light red circles) and the upper kHz QPO (dark blue squares) of 4U~1636\(-\)53 plotted in \autoref{fig:rms_freq_all}.
    Lower panel:  The slope of the best fitting model to the rms amplitude of the upper kHz QPO of 4U~1636\(-\)53 plotted in \autoref{fig:rms_freq_all}
    (a coloured version of this figure is available in the on-line version of the paper)
    }
\end{figure}

In \autoref{fig:rms_freq_all} we show the conditional distribution of the rms amplitude as a function of QPO frequency for both kHz QPOs in different energy bands.
It is apparent from \autoref{fig:rms_freq_all} that the shape of the rms amplitude with frequency remains roughly the same as in \autoref{fig:rms_freq_full} for both kHz QPOs, but the overall amplitude increases as the energy increases from \(4\) keV to \(19\) keV.
As in \autoref{sec:marginals}, we fit the conditional rms-vs-frequency relation at different energies with analytical functions. For the lower kHz QPO we fit a Gaussian function (\autoref{eq:gauss}). The centroid and the width of the Gaussian are consistent with being the same for all energy bands, so we fitted all these  curves simultaneously with these parameters tied during the fit, whereas the normalisation of the Gaussian was left free to vary. The best fit yields a centroid $\nu_0 = 761 \pm 3$ Hz and a width $\sigma = 130 \pm 3$ Hz, with a a reduced chi-squared \(\chi^{2}_{\nu} = 1.8\) for 68 degrees of freedom. We plot the normalisation of the Gaussian as a function of energy in the upper panel of \autoref{fig:normgausian} as red data points.

A look at the right panel of \autoref{fig:rms_freq_all} suggests that the slope of the rms amplitude of the upper kHz QPO becomes steeper as photon energy increases, and that  the hump in the rms vs frequency at $\sim 800$ Hz in \autoref{fig:rms_freq_full} \citep[see also][]{Ribeiro2017b} is present, and at roughly the  same frequency, in all energy bands, the exception being the highest energy band for which we only have upper limits for the rms amplitude.
To quantify this, we fitted  all the curves simultaneously with a model consisting of a linear function that decreases with energy parametrised as in \autoref{eq:uplinear}.
We also added a Gaussian function (\autoref{eq:gauss}) to the model to fit the hump.
We kept the intercept point, $\nu_i$, tied during the fit, since when we let it free to vary between energy bands the best fitting values are consistent with being the same for all the energy bands.
The Gaussian parameters, with exception of the normalisation, $N$, were tied together between all energy bands since, when we left these parameters free the best fitting values were consistent with being the same.
The best fit yields $\nu_i = 1398 \pm 24 $ Hz. For the Gaussian component we obtain $\nu_0 = 771 \pm 16$ Hz and $\sigma = 93 \pm 21$ Hz. This fit yielded a reduced chi-square \(\chi^{2}_{\nu} = 0.6\) for 46 degrees of freedom.
We show the slope of the linear function as a function of energy in the lower panel of \autoref{fig:normgausian} and the normalisation of the hump as dark blue data points in the upper panel of the same figure.

We also fitted the data on both panels of \autoref{fig:rms_freq_all} simultaneously and tied the parameters \(\nu_0\) and \(\sigma\) of the Gaussian function that describes the shape of the relation for the lower kHz QPO and the same parameters of the Gaussian function associated with the \textit{hump} in the relation for the upper kHz QPO.
The fit yielded a chi-square \(\chi^{2} = 153.80\) for 116 degrees of freedom.
The best-fitting parameter for the Gaussian function were \(\nu_{0} = 761 \pm 3\)~Hz and \(\sigma = 129 \pm 4\)~Hz, and, the other parameters were consistent with the ones presented above, including the normalisations of the Gaussian shown in \autoref{fig:normgausian}.
Fitting the data from the lower and upper kHz QPO separately as described in the last two paragraphs yielded a combined chi-square of \(\chi^{2} = 151.4\) for 114 degrees of freedom. These values yield an F-test probability of \(\sim 0.4\), meaning that there is no statistical benefit in keeping the Gaussian parameters of the lower kHz QPO and the hump of the upper kHz QPO free.
In other words, our results indicate that the parameters of the Gaussian that fits the rms-frequency relation of the lower kHz QPO, are consistent with those of the Gaussian that fits the \textit{hump} in the rms-frequency relation of the upper kHz QPO.

\begin{figure*}
    \centering
 	\includegraphics[width=0.9\paperwidth]{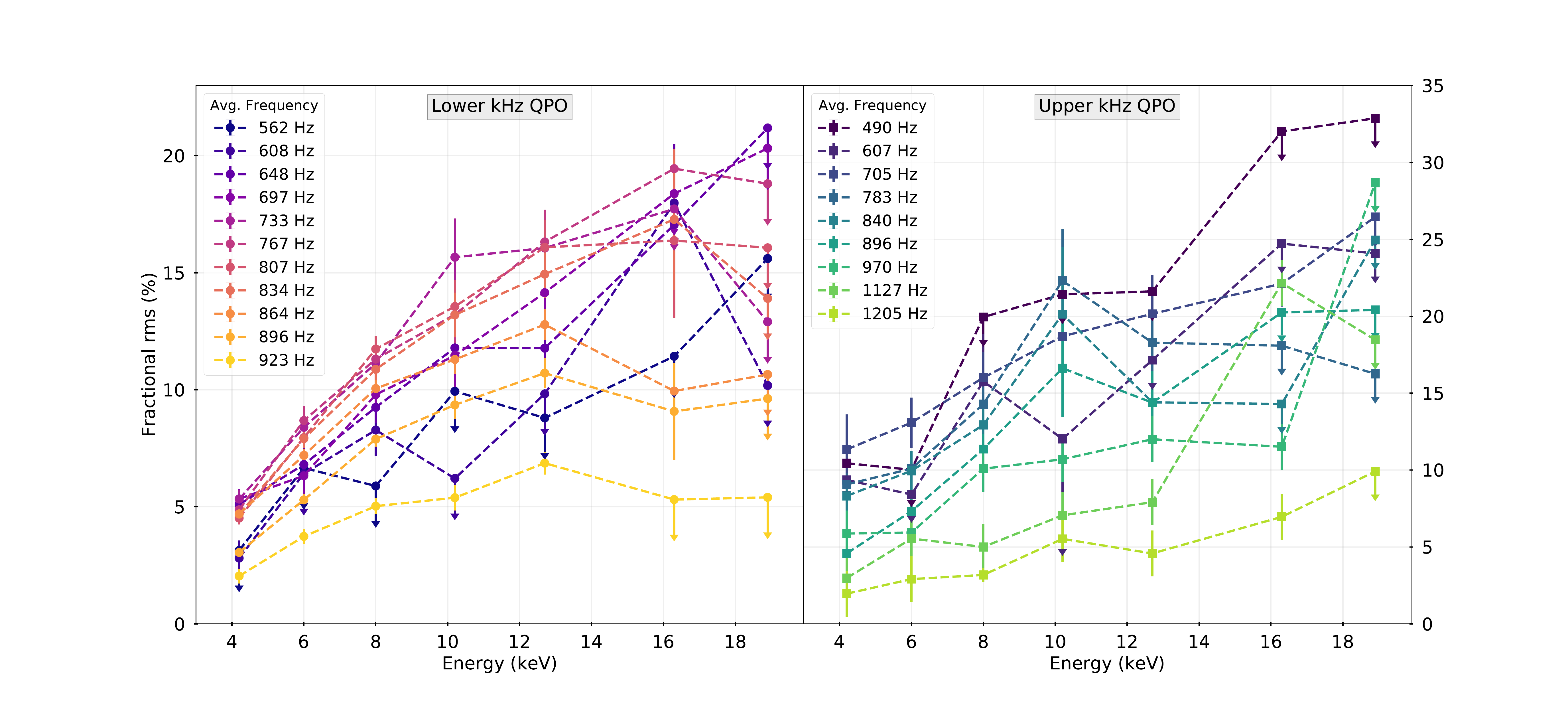}
    \caption{\label{fig:rms_en_all} The conditional distribution of the rms amplitude of the lower and upper kHz QPO, left and right panels, respectively, of 4U~1636\(-\)53 as a function of photon energy for given QPO frequency.
    (A colour version of this figure is available in the on-line version of the paper)
    }
 \end{figure*}

In \autoref{fig:rms_en_all} we show, separately for each QPO, the conditional distributions of the rms amplitude as a function of energy for each QPO frequency interval.
The rms amplitude of both QPOs increases with energy as energy increases from \(\sim 4\)~keV to \(\sim 14\)~keV, above 14~keV the data presents several upper limits and the trend is no longer clear.
The slope of the rms vs. energy relation of the lower kHz QPO (left panel of \autoref{fig:rms_en_all}) appears to increase at first and then decrease as the frequency of the QPO increases further. On the other hand, the slope of the rms vs. energy relation of the upper kHz QPO decreases as the frequency of the QPO increases.

\begin{figure}
    \centering
    \includegraphics[width=0.9\columnwidth]{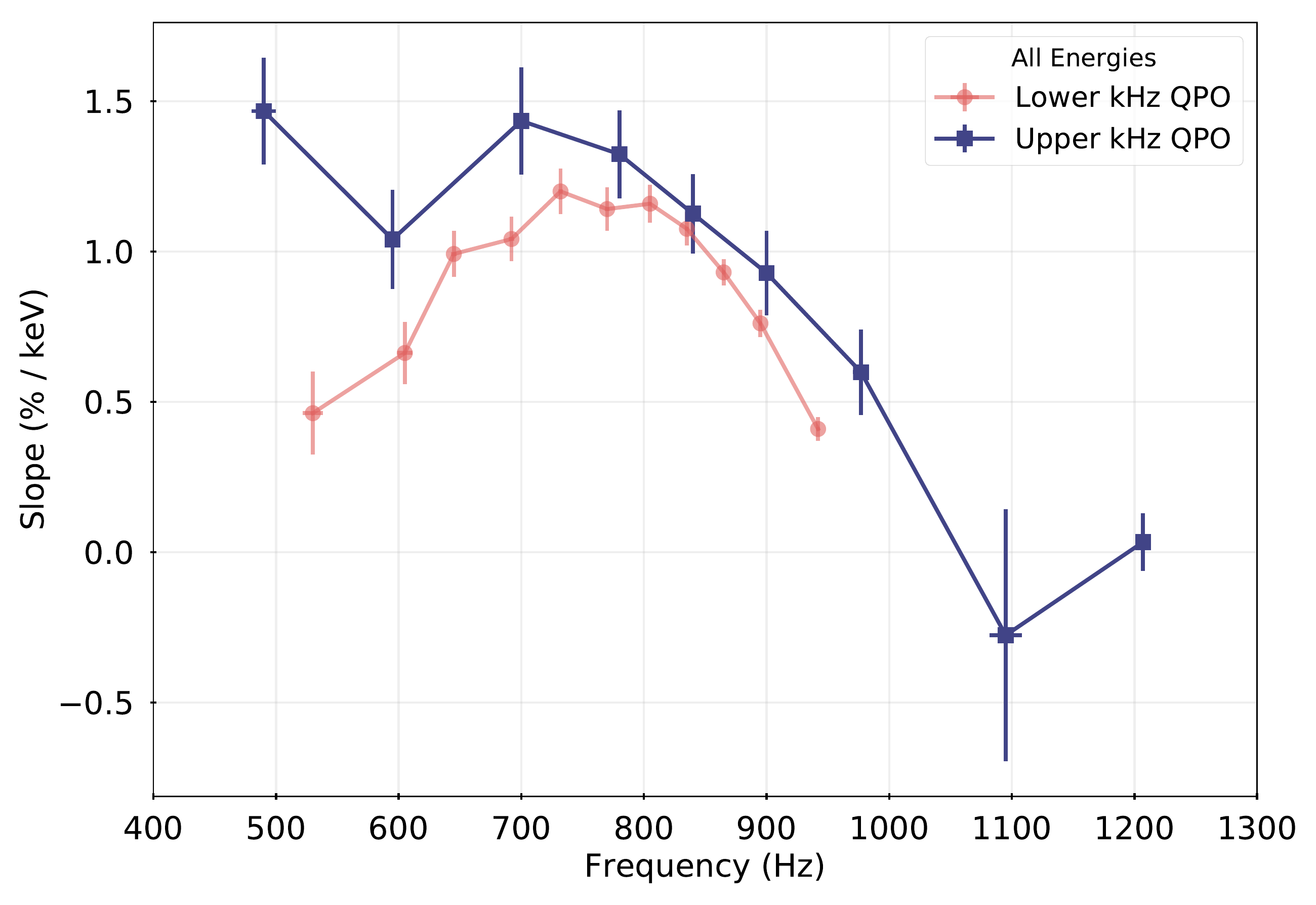}
    \caption{\label{fig:slopes_all} The slope before the break of the conditional distribution of the rms amplitude as a function of energy for given frequency intervals (\autoref{fig:rms_en_all}), for both kHz QPOs. The lower kHz QPO is shown as light red circles and the upper kHz QPO is shown as dark blue squares.
    (A colour version of this figure is available in the on-line version of the paper)
    }
\end{figure}

To quantify this, we fitted the conditional distributions in \autoref{fig:rms_en_all} with a broken line model (\autoref{eq:brokenline}) for both kHz QPOs. At first we let all other parameters free to vary during the fit, but we noticed that the break energy and the slope after the break were consistent with being the same, for each QPO separately in all QPO frequency intervals.
We therefore tied the energy break and the slope above the break to be the same in all QPO frequency intervals and allowed only the slope before the break to vary for the different frequency intervals.
For the lower kHz QPO the best fit yields an energy break $E_{break} = 12.0 \pm 0.2$ keV and a slope after the break $s_2 = -0.5 \pm 0.1$ \%/keV, the reduced chi-squared is \(\chi^{2}_{\nu} = 1.4\) for 64 degrees of freedom. We show the slope before the break, $s_1$, as a function of QPO frequency as light red data points in \autoref{fig:slopes_all}.

For the upper kHz QPO the slope after the break is consistent with being zero in all QPO frequency intervals, so we fixed  $s_2 = 0$. The best fit to the upper kHz QPO yields an energy break $E_{break} = 9.3 \pm 0.7$ keV and a reduced chi-squared \(\chi^{2}_{\nu} = 0.7\) for 52 degrees of freedom; we show the slope before the break as a function of QPO frequency as dark blue data points in \autoref{fig:slopes_all}.

The normalisation of the Gaussian fitted to the curve of the rms amplitude vs. frequency for the lower kHz QPO (\autoref{fig:rms_freq_all}, left panel) reproduces the shape of the rms amplitude vs. energy for that same QPO (\autoref{fig:normgausian}), whereas for the upper kHz QPO the relation is not as straightforward. 
The bottom panel of \autoref{fig:normgausian} shows the slope of the linear function fitted to the rms amplitude vs. frequency relations for the upper kHz.
The slopes increase with energy, but this increase can not be translated into a fractional rms amplitude to allow a direct comparison with the upper panel of \autoref{fig:normgausian}. 
It is interesting to notice that the normalisation of the "hump" in the rms-vs-frequency relation of the upper kHz QPO rms, displays a similar trend to that of the normalisation of the rms-vs-frequency relation of the lower kHz QPO, albeit with smaller amplitudes and larger uncertainties.

In \autoref{fig:rms_en_all} we show the relation between the fractional rms amplitude and photon energy for both kHz QPOs at different QPO frequencies.
The individual relations follow the average behaviour displayed in \autoref{fig:rms_allfreq}. We find that the slope with which the rms increases with photon energy varies with QPO frequency.
The change of slope is reflected as the general trend displayed by the rms amplitude vs. frequency relation in \autoref{fig:rms_freq_full}. 
It is unclear if one is caused by the other, but it is clear that there is a relation between the rms changes with frequency and the slope with which the rms changes with energy, for both kHz QPOs.

\subsection{The joint distribution of the kHz QPOs rms amplitude in the frequency-energy space}\label{sec:2dimensions}

\begin{figure*}
    \centering
    \includegraphics[width=0.75\paperwidth]{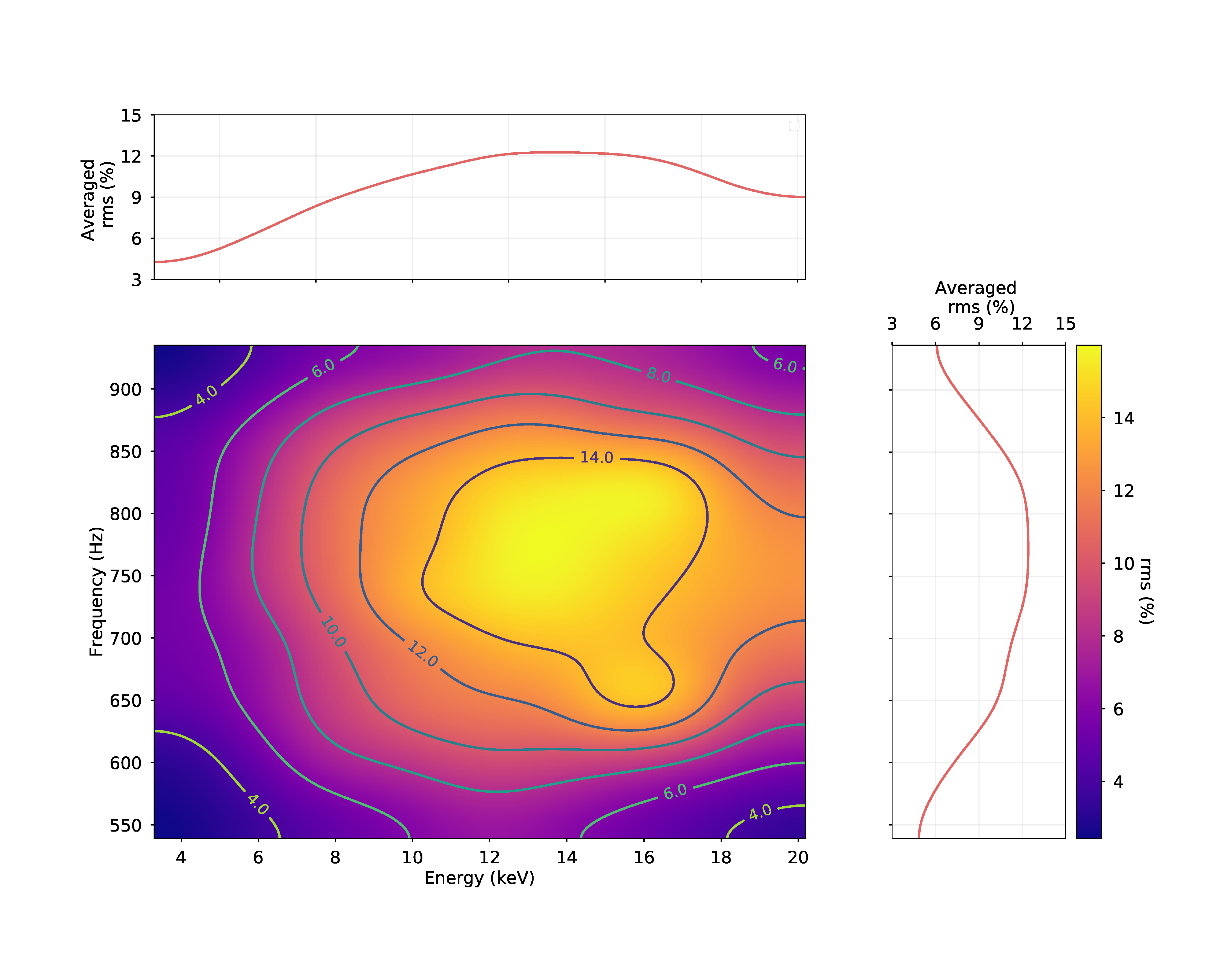}
    \caption{\label{fig:low2d} The joint distribution of the rms amplitude of the lower kHz QPO of 4U~1636$-$53 as a function of photon energy and QPO frequency. The colour scale represents the rms amplitude as indicated in the colour bar at the far right of the Figure. The top and right panels show the marginal distributions averaged over, respectively, QPO frequency and energy. Contour lines are shown on top of the colour map to aid visualisation.
    (A colour version of this figure is available in the on-line version of the paper)
    }
\end{figure*}

\begin{figure*}
   \centering
   \includegraphics[width=0.75\paperwidth]{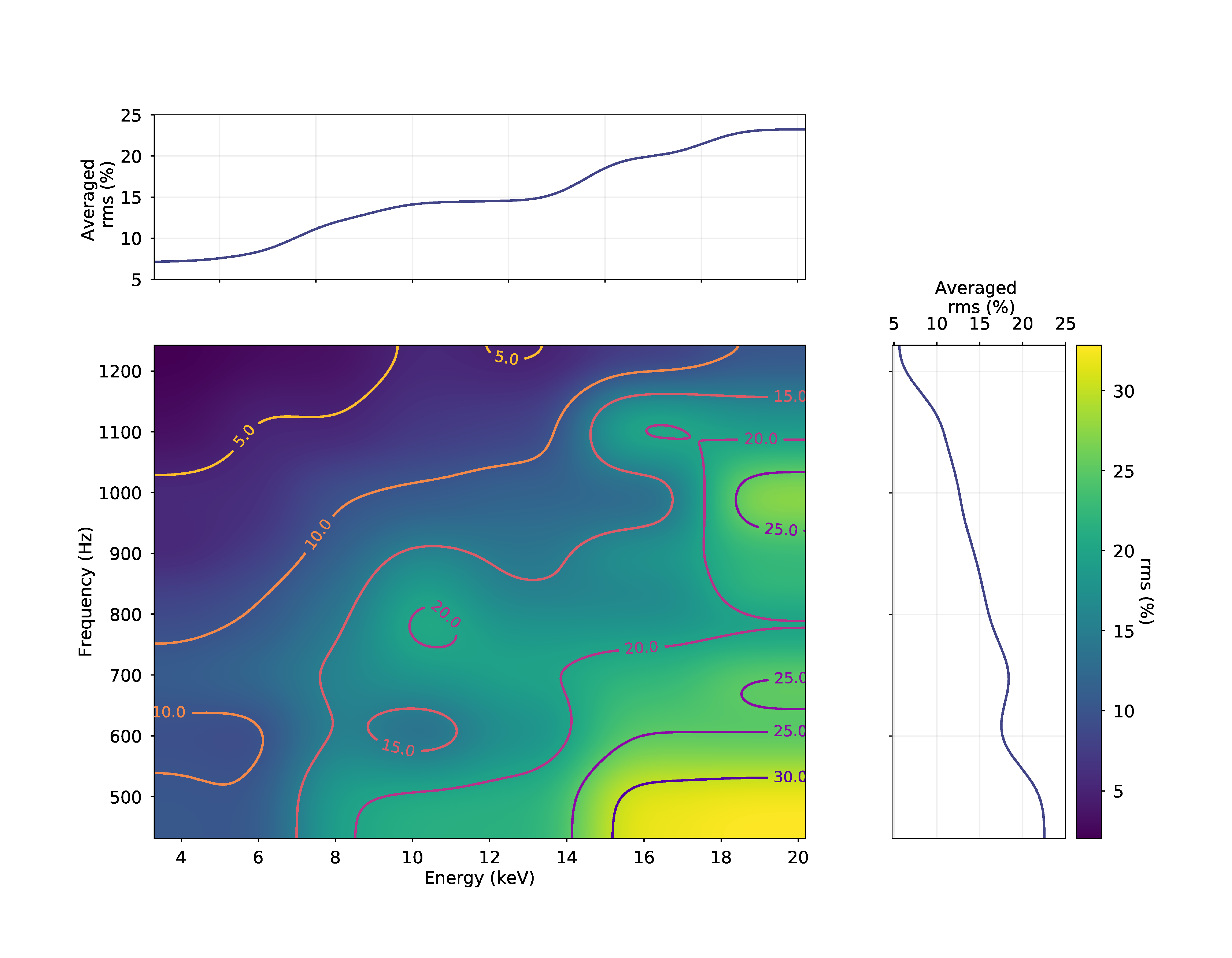}
   \caption{\label{fig:up2d} Same as \autoref{fig:low2d} for the upper kHz QPO of 4U~1636$-$53.
   (A colour version of this figure is available in the on-line version of the paper)
   }
\end{figure*}

Using the conditional distributions from Figures \ref{fig:rms_freq_all} and \ref{fig:rms_en_all}, in Figures \ref{fig:low2d} and \ref{fig:up2d} we plot the rms amplitude of the kHz QPOs as a colour map in the 2 dimensional grid of QPO frequency and energy.
We provide the measured values in \autoref{tab:joint}.
To improve the visualisation we smoothed the data using a 2-dimensional Gaussian kernel.
To represent the upper limits in these 2D plots, we used instead the rms values given by the best-fitting models, described in the previous sub-section, to represent the expected values.

\begin{table*}
  \centering
	\caption{\label{tab:joint}
    The fractional rms amplitude of the kHz QPOs in 4U~1636\(-\)53 as a function of the energy band and the QPO frequency interval.
    Uncertainties are the propagated 1\(\sigma\) error from the best fit to each power spectrum. 
    }
	\begin{tabular}{lccccccc} 
    \toprule
        Frequency & \multicolumn{7}{c}{Energy band (keV)} \\
        interval (Hz) & 3.0--5.0 & 5.0--7.0 & 7.0--9.0 &
                     9.0--11.0 & 11.0--15.0 & 15.0--17.0 & 17.0--20.0 \\
    \midrule
    \multicolumn{4}{l}{Lower kHz QPO} \\
    \midrule
         \(470\)--\(590\) & \(2.0 \pm 1.4\ (3.1\ ^{ *})\) &
                            \(5.0 \pm 1.2\ (6.7\ ^{ *})\) &
                            \(4.1 \pm 1.4\ (5.9\ ^{ *})\) &
                            \(6.6 \pm 2.6\ (9.9\ ^{ *})\) &
                            \(5.9 \pm 2.2\ (8.8\ ^{ *})\) & 
                            \(0.0 \pm 10.2\ (11.4\ ^{ *})\) & 
                            \(0.0 \pm 13.0\ (15.6\ ^{ *})\)\\
         \(590\)--\(620\) & \(2.8 \pm 0.7\) &
                            \(5.0 \pm 0.9\ (6.4\ ^{ *})\) &
                            \(8.3 \pm 1.1\) &
                            \(0.0 \pm 5.2\ (6.2\ ^{ *})\) &
                            \(7.9 \pm 1.4\ (9.8\ ^{ *})\) &
                            \(13.6 \pm 3.8\ (18.0\ ^{ *})\) & 
                            \(0.0 \pm 11.8\ (10.2\ ^{ *})\)\\
         \(620\)--\(670\) & \(5.1 \pm 0.5\) &
                            \(6.8 \pm 0.7\) &
                            \(9.3 \pm 0.7\) &
                            \(11.8 \pm 1.6\) &
                            \(11.8 \pm 1.3\) &
                            \(17.1 \pm 3.5\) & 
                            \(16.5 \pm 5.0\ (21.2\ ^{ *})\)\\
         \(670\)--\(715\) & \(5.3 \pm 0.4\) &
                            \(6.3 \pm 0.8\) &
                            \(9.8 \pm 0.6\) &
                            \(11.5 \pm 1.5\) &
                            \(14.2 \pm 1.3\) &
                            \(14.5 \pm 3.3\ (18.4\ ^{ *})\) & 
                            \(14.8 \pm 5.2\ (20.3\ ^{ *})\)\\
         \(715\)--\(750\) & \(5.3 \pm 0.4\) &
                            \(8.4 \pm 0.7\) &
                            \(11.1 \pm 0.7\) &
                            \(15.7 \pm 1.7\) &
                            \(16.1 \pm 1.4\) &
                            \(14.7 \pm 3.5\ (17.7\ ^{ *})\) & 
                            \(0.0 \pm 10.9\ (12.9\ ^{ *})\)\\
         \(750\)--\(790\) & \(4.9 \pm 0.4\) &
                            \(8.7 \pm 0.6\) &
                            \(11.3 \pm 0.6\) &
                            \(13.2 \pm 1.5\) &
                            \(16.3 \pm 1.4\) &
                            \(14.8 \pm 3.6\ (19.5\ ^{ *})\) & 
                            \(12.3 \pm 5.7\ (18.8\ ^{ *})\)\\
         \(790\)--\(820\) & \(4.5 \pm 0.3\) &
                            \(8.0 \pm 0.5\) &
                            \(11.7 \pm 0.5\) &
                            \(13.6 \pm 1.1\) &
                            \(16.1 \pm 1.2\) &
                            \(16.7 \pm 3.5\) & 
                            \(12.2 \pm 4.0\ (16.1\ ^{ *})\)\\
         \(820\)--\(850\) & \(4.7 \pm 0.2\) &
                            \(7.9 \pm 0.4\) &
                            \(10.9 \pm 0.5\) &
                            \(13.2 \pm 1.0\) &
                            \(14.9 \pm 1.0\) &
                            \(16.8 \pm 2.9\) & 
                            \(10.8 \pm 3.5\ (13.9\ ^{ *})\)\\
         \(850\)--\(880\) & \(4.7 \pm 0.2\) &
                            \(7.2 \pm 0.3\) &
                            \(10.1 \pm 0.4\) &
                            \(11.3 \pm 0.6\) &
                            \(12.8 \pm 0.7\) &
                            \(9.7 \pm 0.9\) & 
                            \(0.0 \pm 19.0\ (10.7\ ^{ *})\)\\
         \(880\)--\(910\) & \(3.1 \pm 0.2\) &
                            \(5.3 \pm 0.3\) &
                            \(7.9 \pm 0.3\) &
                            \(9.4 \pm 0.6\) &
                            \(10.7 \pm 0.6\) &
                            \(10.3 \pm 2.2\) & 
                            \(6.7 \pm 2.0\ (9.6\ ^{ *})\)\\
         \(910\)--\(975\) & \(2.0 \pm 0.3\) &
                            \(3.7 \pm 0.3\) &
                            \(5.0 \pm 0.3\) &
                            \(5.4 \pm 0.5\) &
                            \(6.9 \pm 0.5\) &
                            \(3.5 \pm 1.1\ (5.3\ ^{ *})\) & 
                            \(3.0 \pm 1.1\ (5.4\ ^{ *})\)\\
    \midrule
        \multicolumn{4}{l}{Upper kHz QPO} \\
    \midrule
        \(440\)--\(540\)  & \(10.4 \pm 1.8\) &
                            \(10.0 \pm 2.0\) &
                            \(18.0 \pm 1.6\ (19.9\ ^{ *})\) &
                            \(16.2 \pm 4.0\ (21.4\ ^{ *})\) &
                            \(18.5 \pm 2.5\ (21.6\ ^{ *})\) &
                            \(25.2 \pm 5.8\ (32.0\ ^{ *})\) & 
                            \(23.5 \pm 7.9\ (32.9\ ^{ *})\)\\
        \(540\)--\(650\) &  \(9.4 \pm 1.2\) &
                            \(8.4 \pm 1.4\) &
                            \(15.8 \pm 3.1\) &
                            \(12.0 \pm 7.2\) &
                            \(14.8 \pm 1.9\ (17.2\ ^{ *})\) &
                            \(19.2 \pm 4.7\ (24.7\ ^{ *})\) & 
                            \(15.6 \pm 7.1\ (21.1\ ^{ *})\)\\
        \(650\)--\(750\) &  \(11.3 \pm 2.3\) &
                            \(13.1 \pm 1.6\) &
                            \(16.0 \pm 1.7\) &
                            \(18.7 \pm 4.0\) &
                            \(20.2 \pm 2.6\) &
                            \(15.0 \pm 5.6\ (22.1\ ^{ *})\) & 
                            \(18.0 \pm 7.3\ (26.5\ ^{ *})\)\\
        \(750\)--\(810\) &  \(9.1 \pm 0.7\) &
                            \(10.1 \pm 1.1\) &
                            \(14.3 \pm 1.1\) &
                            \(22.3 \pm 3.4\) &
                            \(18.3 \pm 1.9\) &
                            \(12.2 \pm 4.8\ (18.1\ ^{ *})\) & 
                            \(1.7 \pm 45.1\ (16.3\ ^{ *})\)\\
        \(810\)--\(870\) &  \(8.3 \pm 0.9\) &
                            \(10.0 \pm 1.3\) &
                            \(12.9 \pm 1.1\) &
                            \(15.2 \pm 1.9\) &
                            \(15.3 \pm 1.5\) &
                            \(4.0 \pm 14.2\ (14.3\ ^{ *})\) & 
                            \(18.5 \pm 6.0\ (25.0\ ^{ *})\)\\
        \(870\)--\(930\) &  \(4.6 \pm 2.3\) &
                            \(7.3 \pm 1.0\) &
                            \(11.4 \pm 1.0\) &
                            \(16.6 \pm 3.2\) &
                            \(14.4 \pm 2.1\) &
                            \(14.9 \pm 4.4\ (20.2\ ^{ *})\) & 
                            \(10.7 \pm 8.9\ (20.4\ ^{ *})\)\\
        \(930\)--\(1025\) & \(4.5 \pm 0.9\ (5.9\ ^{ *})\) &
                            \(3.1 \pm 2.6\ (5.9\ ^{ *})\) &
                            \(8.8 \pm 0.8\ (10.1\ ^{ *})\) &
                            \(6.7 \pm 3.2\ (10.7\ ^{ *})\) &
                            \(9.5 \pm 1.8\ (12.0\ ^{ *})\) &
                            \(0.0 \pm 11.6\ (11.5\ ^{ *})\) & 
                            \(21.1 \pm 6.5\ (28.7\ ^{ *}\)\\
        \(1025\)--\(1165\)& \(0.0 \pm 3.2\ (3.0\ ^{ *})\) &
                            \(0.0 \pm 4.7\ (5.5\ ^{ *})\) &
                            \(0.0 \pm 4.6\ (5.0\ ^{ *})\) &
                            \(0.0 \pm 8.0\ (7.1\ ^{ *})\) &
                            \(0.0 \pm 7.2\ (7.9\ ^{ *})\) &
                            \(8.8 \pm 14.3\ (22.2\ ^{ *})\) & 
                            \(0.0 \pm 21.5\ (18.5\ ^{ *})\)\\
        \(1165\)--\(1250\)& \(1.3 \pm 0.5\ (2.0\ ^{ *})\) &
                            \(2.2 \pm 0.5\ (2.9\ ^{ *})\) &
                            \(3.2 \pm 0.5\) &
                            \(4.5 \pm 0.7\ (5.5\ ^{ *})\) &
                            \(3.4 \pm 0.8\ (4.6\ ^{ *})\) &
                            \(2.2 \pm 6.0\ (7.0\ ^{ *})\) & 
                            \(5.3 \pm 4.1\ (9.9\ ^{ *})\)\\\\
    \bottomrule
   		\multicolumn{4}{l}{\(^{*}\) Upper limit at \(95 \% \) confidence. }
	\end{tabular}
\end{table*}

We show the rms amplitude in the frequency-and-energy space for the lower kHz QPO in \autoref{fig:low2d}; the top panel shows the marginal distribution of the rms amplitude as a function of energy averaged over frequency, and the right panel shows the marginal distribution of the rms amplitude as a function of frequency averaged over energy.
As expected, the marginal distributions are a smoothed version of the plots in Figures \ref{fig:rms_freq_full} and \ref{fig:rms_allfreq}.
We over plot contour lines to help visualise the structure of the distribution. We show the same results for the upper kHz QPO in \autoref{fig:up2d}. 

We see in \autoref{fig:low2d} that the rms of the lower kHz QPO is almost symmetric in the frequency axis, consistent with the fact that the marginal and conditionals distribution of the rms amplitude vs. QPO frequency can be fitted with a Gaussian function.
As a function of energy we can see that the rms amplitude of the lower kHz QPO increases and then decreases with energy.

The topology of the rms amplitude of the upper kHz QPO shows a more complex structure, in part probably due to the fact that for several energy bands and frequency intervals we only have upper limits to the rms amplitude. We can see in \autoref{fig:up2d}, however, that the rms amplitude increases as photon energy increases while it simultaneously decreases as the QPO frequency increases.

\section{Discussion}\label{sec:disc}

We present the first study of the distribution of the fractional rms amplitude of the kHz QPOs in a NS-LMXB both as a function of photon energy and QPO frequency.
Previous studies of the kHz QPO had examined the dependence of the rms amplitude upon either energy or frequency, marginalising the dependence of the rms amplitude upon the other quantity.
We find that in 4U~1636\(-\)53 the change of the rms amplitude of both kHz QPOs with QPO frequency depends on energy, and is in fact connected to changes of the slope in the rms energy spectrum (rms amplitude vs. energy) of the QPO.
In the case of the lower kHz QPO the slope in the rms energy spectrum increases as the QPO frequency increases from \(\sim500\)~Hz to \(\sim750\)~Hz, and then decreases when the QPO frequency increases further from \(\sim 750\)~Hz to \(\sim 950\)~Hz.
In parallel with that, the 2\(-\)60~keV rms amplitude of the lower kHz QPO follows the same behaviour with QPO frequency.
In the case of the upper kHz QPO the slope in the rms spectrum generally decreases as the QPO frequency increases from \(\sim 500\)~Hz up to \(\sim1200\)~Hz, showing a local hump at around \(~700\)~Hz. Also in this case, the frequency dependence of the 2\(-\)60~keV rms amplitude of the upper kHz QPO mirrors this behaviour.
Finally, we find that the kHz QPO frequency is the same at different energy bands, and that the rms amplitude of the lower kHz QPO in 4U~1636\(-\)53 drops significantly at energies above $12$~keV, compared to the extrapolation of the rms-energy relation below that energy.

\subsection{The drop of the rms amplitude of the lower kHz QPO at high energies}
\label{sec:disc-drop}

The shape of the rms energy spectrum is essential to understand the origin of kHz QPOs, since it provides a link between a timing property (amplitude of the QPO) and a spectral property (high energy emission) of the source.
Above 10~keV, where the kHz QPOs have an rms amplitude of \(\sim 10\)~\% or more \citep[\autoref{fig:rms_allfreq} and \autoref{fig:rms_en_all}, see also][]{Berger1996a, Gilfanov2003, Mendez2001, Peille2015, Troyer2018} the accretion disc contribution to the energy spectrum of LMXBs is negligible, and the spectrum is dominated by the Comptonising component.
We find that the rms amplitude of the lower kHz QPO declines significantly at photon energies above \(\sim 12\)~keV, implying that the efficiency of the radiative mechanism responsible for this QPO decreases as the energy increases above that energy.

\cite{Mukherjee2012} reported that the rms fractional amplitude of the lower kHz QPO in 4U~1728\(-\)34 drops at high energies. However, the result reported by \cite{Mukherjee2012} is based on an upper limit measurement of the fractional rms amplitude of the kHz QPO in this source at a high energy band (approximately 10--20 keV) that was calculated in such a manner that underestimates the real QPO amplitude. \cite{Mukherjee2012} calculated their upper limit from the power measured in a single frequency bin of width \(\sim3 \)~Hz of the power spectrum at the expected QPO frequency; this calculation ignores the width of the QPO which was around 10 Hz when the QPO was significantly detected in the power spectrum \citep{Mukherjee2012}. Considering the width of the QPO, the correct 95~\% confidence upper limit would be \(\sim 8\)~\%. This revised upper limit is consistent with the results of \cite{Peille2015} and \cite{Troyer2018}, who find that the rms amplitude of the lower kHz QPO in 4U~1728\(-\)34 in the 10--20 keV energy band is \(\sim 8\)--\(9\)~\%

While our results for 4U~1636\(-\)53 agree with their claim that in 4U~1728\(-\)34 the fractional rms amplitude of the lower kHz QPO drops at high energies, the drop is not as abrupt as the one reported by \cite{Mukherjee2012} for 4U~1728\(-\)34  on the basis of an underestimated upper limit.

\subsection{The radiative mechanism(s) behind the kHz QPOs}
\label{sec:disc-mech}

The plot of the rms amplitude of the upper kHz QPO as a function of frequency for the full energy band displays a local maximum at \(\sim765\)~Hz that we call a "hump".
As discussed by \cite{Ribeiro2017b}, it is curious this "hump" happens at a frequency with the maximum in the relation between the rms amplitude and the frequency of the lower kHz QPO. 
From our fits we can not discard that the hump is present in all energy bands investigated (see \autoref{fig:rms_freq_all}).
Our results are consistent with the scenario proposed by \cite{Ribeiro2017b} \citep[see also ][]{Mendez2006, Sanna2010, DeAvellar2013} that there is a radiative mechanism that is more efficient at \(\sim750\)~Hz and influences the rms amplitude of both kHz QPO, whereas there is another mechanism that acts only upon the upper kHz QPO and drives the more or less linear decay of the rms amplitude of this QPO with frequency.

\cite{Lee2001} \citep[see also ][]{DeAvellar2013, Zhang2016, Ribeiro2017b} suggested that this mechanism that acts both upon the lower and the upper kHz QPO is a resonance between the source of soft photons (the neutron-star or the accretion disc) and the Comptonising medium.
The quality factor of the lower kHz QPO peaks at around the same frequency \citep{Belloni2005, Barret2006a}, and this is also the frequency interval that shows the largest phase lags and the highest coherence between low- and high-energy signals of the lower kHz QPO~\citep{DeAvellar2013, DeAvellar2016}.

As shown on \autoref{fig:rms_allfreq}, the fractional rms amplitude of the upper kHz QPO increases with photon energy.
A fit with a broken-line model indicates that the rms amplitude either remains constant or continues increasing as energy increases; given the limitations of the RXTE/PCA instrument we can not strongly argue for one or the other scenario. 
It is interesting to notice that the rms energy spectrum of high-frequency QPOs in black-hole LMXBs \citep[e.g., GRS~1915\(+\)105][]{Belloni2013} shows a similar trend to the marginal distribution of rms vs. energy of the upper kHz QPO in 4U~1636\(-\)53; a deeper look into how this behaviour changes with QPO frequency could shed light on the connection between the variability of neutron-star and black-hole LMXBs.

\subsection{Implications of the conditional distributions of rms amplitude of the kHz QPO}
\label{sec:disc-cond}

By analysing the relation between fractional rms amplitude with frequency, we show (\autoref{fig:rms_freq_all}) that this relation of the lower kHz QPO maintains its shape in different energy bands, and only its normalisation changes.
The slope of the relation of the rms amplitude of the upper kHz QPO with frequency, however, shows an increase and not just an achromatic shift.

As shown in \autoref{fig:slopes_all} the slope of the lower kHz QPO rms spectrum first increases and then decreases with frequency while, at the same time, the temperature of the accretion disc only increases \citep{Sanna2013a, Lyu2014}, suggesting that the slope of the relation between the rms fractional amplitude and energy for the lower kHz QPO is not driven by the soft emission from the disc.
The slope in the rms vs.\ energy relation (\autoref{fig:rms_en_all}) changes with QPO frequency in a similar fashion as the rms changes with QPO frequency, and shows a seemingly preferable frequency around \(\sim 750\) Hz for both the upper and lower kHz QPO.
This reinforces the idea of a common mechanism between the two QPOs, but with an extra component for the upper kHz QPO as mentioned on \autoref{sec:disc-mech}.

The dependence of the fractional rms amplitude of the kHz QPOs upon energy presented here can be compared to that of the type-C QPOs at \(\sim 0.15\)--\(12\)~Hz in black-holes LMXBs presented by \cite{Zhang2017} and \cite{Huang2018}.
Here we find that the rms amplitude vs energy relation increases and then decreases with energy, for the lower kHz QPO, while the rms amplitude only increases with energy for the upper kHz QPO and the slope of the rms amplitude vs energy relation for the upper kHz QPO decreases as the QPO frequency increases. Both \cite{Zhang2017} and \cite{Huang2018} find that the slope of the rms vs energy relation for the type-C QPO  in, respectively, GX~339\(-\)4 and MAXI~J1535\(-\)571 increases with QPO frequency.

For the upper kHz QPO, the best fitting parameters for the relation in \autoref{fig:rms_freq_all} yield \(\nu_i = 1398 \pm 24\)~Hz as the frequency at which the rms amplitude goes to zero, and, as mentioned in \autoref{sec:conditionals}, this value is consistent with being the same across the different energy bands. The fact that the rms of the upper kHz QPO goes to zero at a specific frequency could reflect the dynamical mechanism that produces the QPO, e.g. if the inner radius of the accretion disk reaches the innermost stable circular orbit when the upper kHz QPO reaches that frequency.

In the sonic-point model proposed by \cite{Miller1998}, this maximum QPO frequency can be used to get an upper limit of the mass of the neutron-star \citep{Kluzniak1990, VanDoesburgh2018}:

\begin{equation}\label{eq:mass}
    M_{NS} \leq 2.2 \cdot (1000/\nu_{QPO}^{max}) \cdot (1 + 0.75j) \cdot M_{\odot}\ , \\
\end{equation}

\noindent where $j$ is the angular momentum of the neutron-star, which depends on the choice of equation of state~\citep[see][]{Morsink2002}. For the minimum and maximum values of $j$, given by \cite{Morsink2002}, and for \(\nu_{QPO}^{max} = \nu_i\), we get upper limits for  the mass of the neutron-star in 4U~1636\(-\)53 \(M_{NS} \leq 1.77 M_{\odot}\) for $j=0.17$ and \(M_{NS} \leq 2.19 M_{\odot}\) for $j=0.52$. As pointed out by \cite{VanDoesburgh2018}, Equation \ref{eq:mass} does not take into account the oblateness and internal structure of the NS on its rotation rate for fast spinning stars such as 4U~1636$-$53 and numerical methods are necessary to obtain accurate limits on mass
and radius in these cases. The limits given above here are likely underestimated. It is outside of the scope of this paper to discuss the different equations of state or the neutron star oblateness and their implications.

\subsection{The break energy}\label{sec:disc-break}

The rms amplitude as a function of energy of both the lower and upper kHz QPO display a break at around \(E_b \approx 12\)~keV.
This connection between the kHz QPOs and the high energy photons is an important clue to the origin of these oscillations.
The maximum fitted temperature to the Comptonising medium ($kT_e$) in the energy spectrum of 4U~1636\(-\)53 is also around \(\sim 12\)~keV \citep{Ribeiro2017b}, but the fact that the break in the rms amplitude vs. energy relation is consistent with being the same for all QPO frequencies whereas the electron temperature of the corona changes as the QPO frequency changes \citep{Ribeiro2017b} makes it hard to establish a clear connection between $E_b$ and $k$T$_e$.
The presence of the break seems to be important, however, for the modelling of the radiative mechanism that produces the lower kHz QPO, as we discuss next.

\subsection{The kHz QPOs as oscillations of the Comptonised flux}
\label{sec:disc-model}

We used the model of \citeauthor{Kumar2014} (\citeyear{Kumar2014}; see also \citealt{Lee2001}) to describe the relation of the rms amplitude with energy for the lower kHz QPO.
The model of \cite{Kumar2014} has as parameters:
the amplitude of the oscillation of the external heating rate, \(\delta H_{ext}\), where the external heating rate is necessary to balance the cooling effect of inverse Compton scattering, the size of the Comptonising medium, $L$, a parameter that describes the fraction of scattered photons that return to the source of seed photons, or feedback parameter,\(\eta\), the temperature of the seed photon source, $kT_s$, the electron temperature of the Comptonising medium, $kT_e$, and the optical depth of the Comptonising medium, \(\tau\).
Given that our rms spectra has only 7 data points and the model requires 6 parameters, we also used the time-lag spectra presented by \cite{DeAvellar2013}, given that the model predicts both the rms and the lag spectrum. Furthermore, since the model can not reproduce the break in the rms spectrum at high energies \citep[see][for a discussion about the possible reasons for this]{Kumar2014}, we ignored the 2 last energy bins in our spectra during the fits.

Since the implementation of a fitting method for the model is computationally expensive, we leave that, together with a detailed exploration of the parameter space, for a subsequent paper (Karpouzas et al. in prep.); here, instead, we tested a set of model parameters and provide an approximate fit to both the rms and lag spectra simultaneously. According to \cite{Kumar2016} the model is degenerate, and one can qualitatively fit the data using either a hot- or a cold-seed photon source. We find the same in this case, as shown in Fig. 10: This Figure shows a hot-seed model with \(L = 1.18\)~km, \(\delta H_{ext} = 0.08\), \(\eta = 0.9\), \(kT_e = 4.6\)~keV, \(kT_s = 1.3\)~keV, and \(\tau = 2.7\), and a cold-seed model with \(L = 4.3\)~km, \(\delta H_{ext} = 0.09\) , \(\eta = 0.6\), \(kT_e = 3.8\)~keV , \(kT_s = 0.4\)~keV, and \(\tau = 10.4\).

\begin{figure}
    \centering
    \includegraphics[width=0.95\columnwidth]{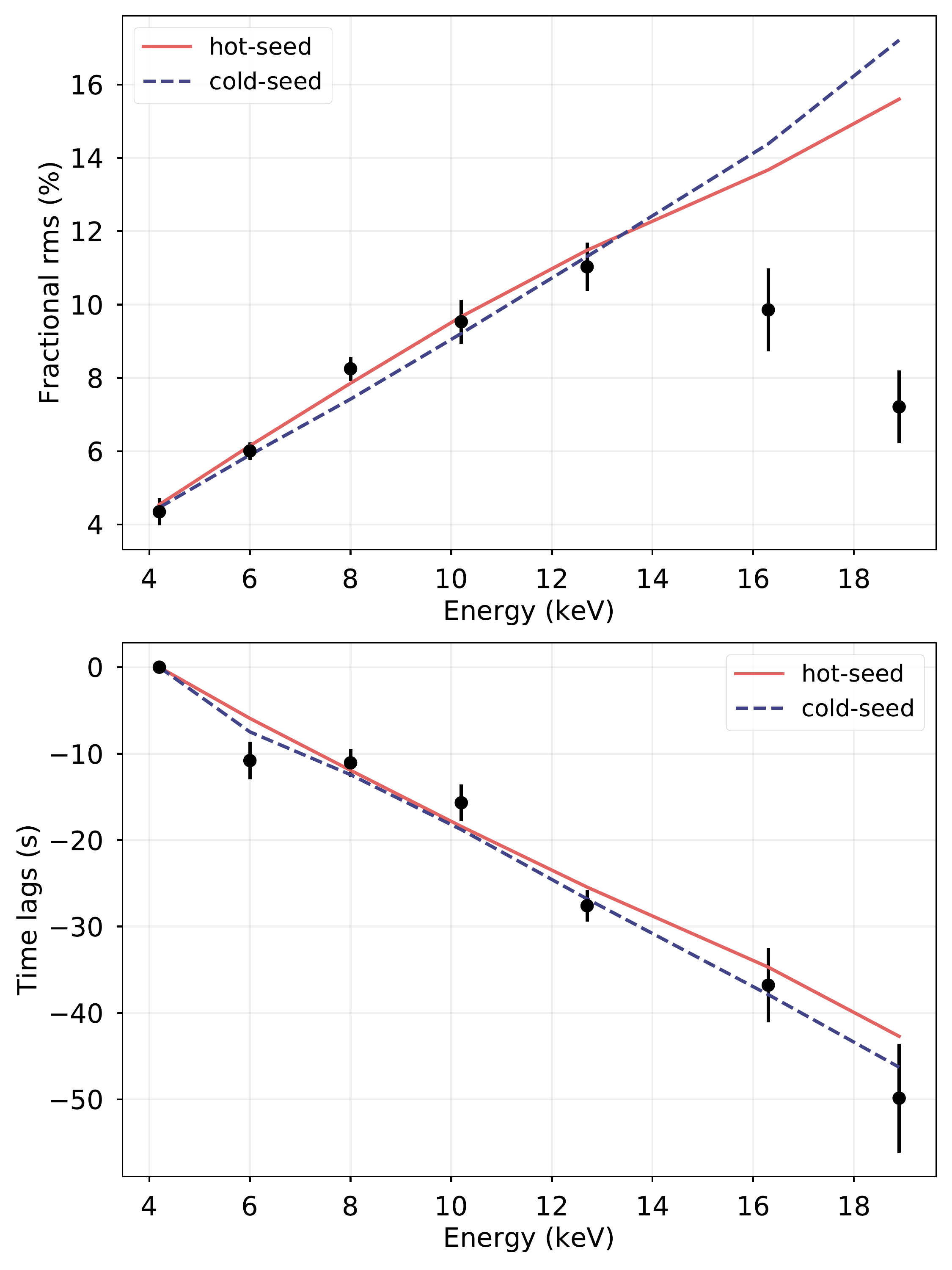}
    \caption{\label{fig:model} Modelling of the rms and time-lag spectra using the the model of Kumar \& Misra (2014).
    The solid red line represents the hot seed photons model, and the dashed blue line represents the cold seed photons model.
    (A colour version of this figure is available in the on-line version of the paper)}
 \end{figure}

\section{Conclusion}

The kHz QPOs in LMXB are likely a result of the interplay between different components in the accretion flow in these systems. We presented a study of the rms amplitude of the kHz QPOs in 4U~1636\(-\)53 in the frequency and energy domain, using the full set of archival observations of this source with the RXTE satellite.
We find that the frequency of the kHz QPOs does not depend on photon energy, whereas its amplitude does, reinforcing the idea that there are at least two distinctive mechanism responsible for the QPOs: a dynamical mechanism that sets the frequency and a radiative mechanism that sets the amplitude. We also show that the rms amplitude of the lower kHz QPO of 4U~1636\(-\)53 drops at high energies which gives us a clue about the nature of the radiative mechanism behind this QPO.

We showed, for the first time for any source of kHz QPOs, the relation of rms amplitude of the kHz QPOs vs. QPO frequency as a function of photon energy and QPO frequency.
Our results give a more complete picture of these oscillations in the spectral-timing domain.
This work helps to pave the way for current and future instruments such as NICER, ASTROSAT, HXMT, eXTP and Athena, the development of new techniques and tools for spectral-timing analysis \citep[e.g.][]{Huppenkothen2019} and the developments of new theoretical models to shed new light upon the nature of the kHz QPOs.

\section*{Acknowledgements}
E.M.R acknowledges the support from Conselho Nacional de Desenvolvimento Cient\'{\i}fico e Tecnol\'{o}gico (CNPq --- Brazil).
MGBA acknowledges CNPq project 150999/2018-6 and FAPESP Thematic Project 2013/26258-4.
G.B. acknowledges funding support from the National Natural Science Foundation of China (NSFC) under grant numbers U1838116 and the CAS Pioneer Hundred Talent Program Y7CZ181002.
This research has made use of data obtained from the High Energy Astrophysics Science Archive Research Center, provided by NASA's Goddard Space Flight Center

\bibliographystyle{mnras}
\bibliography{references} 


\bsp
\label{lastpage}
\end{document}